\documentstyle[twoside,12pt,epsf,aps]{revtex}
\relax
\newcommand{\MET}{\mbox{$\protect \raisebox{.3ex}{$\not$}\et$}}

\def\W{{\em W\/ }}
\def\Z0{${\em Z^0\/}$}

\def\r#1 {$^{#1}$}

\newcommand{\et}{{\rm E}_{\scriptscriptstyle\rm T}}

\def\gepsfcentered#1{
  \def\testit{#1}
  \def\lbracket{[}
  \ifx\testit\lbracket
    \let\dofilecmd=\gepsfwithopt
  \else
    \let\dofilecmd=\gepsfnoopt
  \fi
  \dofilecmd}

\def\gepsfnoopt#1{
  \begin{center}
  \leavevmode
  \epsffile{#1}
  \end{center}}

\def\gepsfwithopt#1 #2 #3 #4]#5{
  \begin{center}
  \leavevmode
  \gepsfmaxx=0.94\textwidth
  \epsffile[#1 #2 #3 #4]{#5}
  \end{center}}

\newdimen\gepsfmaxx
\def\epsfsize#1#2{
  \ifnum \epsfxsize=0
    \ifnum \epsfysize=0
      \ifnum #1 > \gepsfmaxx
        \gepsfmaxx
      \else
        #1
      \fi
    \else
      \epsfxsize
    \fi
  \else
    \epsfxsize
  \fi
}

\begin{document}
\bibliographystyle{unsrt}
\title{Additional studies of the probability that the events
 with a superjet observed by CDF are consistent with the SM
 prediction}
\maketitle
\font\eightit=cmti8
\def\r#1{\ignorespaces $^{#1}$}
\hfilneg
\begin{sloppypar}
\noindent
G.~Apollinari,\r {2}
 M.~Barone,\r {4}  
D.~Benjamin,\r{1}
W.~Carithers,\r {6} 
T.~Dorigo,\r {7} 
I.~Fiori,\r {7} 
M.~Franklin,\r {5}
P.~Giromini,\r {4}
 F.~Happacher,\r {4} 
J.~Konigsberg,\r {3}
M.~Kruse,\r {1} 
S.~Miscetti,\r {4} 
A.~Parri,\r {4}
 F.~Ptohos,\r {4}
and G.~Velev\r {2}
\end{sloppypar}
\vskip .026in
\begin{center}
\r {1}  {\eightit Duke University, Durham, North Carolina  27708} \\
\r {2}  {\eightit Fermi National Accelerator Laboratory, Batavia, Illinois 
60510} \\
\r {3} {\eightit University of Florida, Gainesville, Florida  32611} \\
\r {4} {\eightit Laboratori Nazionali di Frascati, Istituto Nazionale di Fisica
               Nucleare, I-00044 Frascati, Italy} \\
\r {5} {\eightit Harvard University, Cambridge, Massachusetts 02138} \\
\r {6} {\eightit Ernest Orlando Lawrence Berkeley National Laboratory, 
Berkeley, California 94720} \\
\r {7} {\eightit Universita di Padova, Istituto Nazionale di Fisica 
          Nucleare, Sezione di Padova, I-35131 Padova, Italy} \\
\end{center}

\vspace{0.2em}
\begin{abstract}
{In the $\W+$ 2,3 jet data collected by CDF during the 1992-1995 Fermilab 
 collider run, 13 events were observed to contain a superjet when 
 $4.4 \pm 0.6$ events are expected. A previous article detailed the selection
 and the kinematical  properties of these events. The present paper
 provides estimates of the probability that the kinematics of these 13 events 
 is statistically consistent with the standard model prediction.}
 \\
PACS number(s): 13.85.Qk, 13.38.Be, 13.20.He
\end{abstract}
\section{ Introduction}
\label{s-intro}
 The CDF experiment has reported~\cite{anomal}  an excess
 of events in the $\W+2$ and $\W+3$ jet topologies in which the presumed
 heavy-flavor jet contains a soft lepton (SLT tag) in addition to a secondary 
 vertex (SECVTX tag)\footnote{Such a double tag is called supertag in 
 Ref.~\cite{anomal}; jets with a supertag are referred to as superjets.}.
 The rate of these events (13 observed) is larger than what is predicted by a
 simulation of  known standard model (SM) processes (4.4 $\pm$ 0.6 events 
 expected, including single and pair production of top quarks). Various
 kinematical distributions of these events are compared in Ref.~\cite{anomal}
 to what is expected if the excess were simply due to a  statistical 
 fluctuation of the SM contributions. The simulation is cross-checked by 
 comparing to a complementary sample of 42 $\W+2$ and $\W+3$ jet events with 
 SECVTX tags but no supertags.  According to the simulation~\cite{anomal},
 events with a superjet and the complementary data set have quite similar 
 heavy flavor composition. A set of 18 kinematical variables was chosen
 {\em a priori} to look for differences between data and simulation.
 Each data distribution is compared to the SM~expectation using a 
 Kolmogorov-Smirnov (K-S) test~\cite{kuiper,sadoulet}. The probability $P$
 that each distribution is consistent with the SM simulation is derived with
 Monte Carlo pseudo-experiments which include Poisson fluctuations and 
 Gaussian uncertainties in the prediction of each standard model contribution.

 In Ref.~\cite{anomal}, a subset of 9 kinematical variables is selected
 {\em a posteriori} to illustrate the main differences between the data and 
 the simulation: $E_T^{l}$ and $\eta^l$, the transverse energy and
 pseudo-rapidity  of the primary lepton ($l$); $E_T^{suj}$ and $\eta^{suj}$,
 the transverse energy and pseudo-rapidity  of the superjet ($suj$);
 $E_T^{b}$ and $\eta^{b}$, the transverse energy and pseudo-rapidity of the
 additional jets ($b$) in the event; $E_{T}^{l+b+suj}$ and $y^{l+b+suj}$, the
 transverse energy and
 rapidity  of the system $l+b+suj$; and  $\delta \phi^{l,b+suj}$, the
 azimuthal angle between between the primary lepton and the system
 $b+suj$ composed by the superjet and the other jets in the events.
 The first 8 variables test
 if the production cross sections
 ${\displaystyle \frac {d^2 \sigma} {dp_T d\eta} }$
 of each object in the final state is consistent with the
 SM simulation and the ninth variable tests  if the
 data are  consistent with the production and decay of $\W$ bosons from known
 sources.  Table~\ref{tab:tab_comb.1} summarizes the probabilities of these
 comparisons. The SM simulation models correctly the complementary sample of
 data, but has a systematically low probability of being consistent with the
 kinematic distributions of the events with a superjet.
 The use of this subset of variables is well motivated by the fact that it
 provides a simple way to describe in full the kinematics of the final state 
 with relatively modest correlations. However, it is not the only possible 
 choice.

 Table~\ref{tab:tab_comb.2} lists the result of the K-S test of the other 9
 kinematical distributions inspected:  $\MET$, the corrected transverse 
 missing energy; $M_{T}^{W}$, the $\W$ transverse mass calculated using the
 primary lepton and $\MET$; $M^{b+suj}$, $y^{b+suj}$, and $E_T^{b+suj}$,
 the invariant mass, rapidity, and transverse energy of the system $b+suj$
 respectively; $M^{l+b+suj}$, the invariant mass of the system $l+b+suj$;
 $\delta \theta^{b,suj}$ and $\delta \phi^{b,suj}$, the angle and the 
 azimuthal angle between the superjet and the $b$-jets, respectively;
 and $\delta \theta^{l,b+suj}$, the angle between the primary lepton and the
 system $b+suj$. The simulation models correctly these distributions for the
 complementary sample. The probabilities for the events with a superjet are
 systematically lower, but the disagreement between data and simulation is
 much reduced for this second set of variables. 
 This second set of 9 distributions would have been better suited to find
 differences if, for example, events with a superjet were produced by
 the two-body decay of a massive object produced in association with a $\W$ 
 boson or by the three-body decay of a massive object produced in association 
 with large $\MET$.
 
 In Sect.~\ref{sec:eval}, we first evaluate the combined probability that the
 data are statistically consistent with the simulation using different methods
 in order to estimate the effect of possible correlations between kinematic 
 variables. We then study the effect of the bias introduced by the choice of 
 particular sets of kinematical variables which were not motivated by a 
 specific model or by the analysis of an independent data sample.
 Section~\ref{sec:s-concl} summarizes our conclusions.
\newpage
\begin{table}[p]
\begin{center}
\def\arraystretch{0.8}
\caption[]{Results of the K-S comparison between data and simulation
for the first set of 9 kinematical variables.
           $P$ is the probability of making an observation with a
 K-S distance   no smaller than that of the data.}
\begin{tabular}{l c   c}
       & \multicolumn{1}{c}{ Events with a superjet} &
 \multicolumn{1}{c}{ Complementary sample} \\
 Variable      & $P$ (\%)    & $P$ (\%)  \\
  $E_T^{l}$      &  2.6   & 70.9 \\
  $\eta^{l}$    &  0.10    & 72.7 \\
  $E_T^{suj}$      &  11.1     & 43.0 \\
  $\eta^{suj}$       &15.2     & 73.4 \\
  $E_T^{b}$      & 6.7      & 8.6 \\
  $\eta^{b}$    &  6.8      & 80.0 \\
$E_T^{l+b+suj}$    &2.5       &18.8 \\
$y^{l+b+suj}$       &13.8  & 7.8 \\
$\delta \phi^{l,b+suj}$  & 1.0   & 77.9 \\
\end{tabular}             
\label{tab:tab_comb.1}
\end{center}
\end{table}

\begin{table}[p]
\begin{center}
\def\arraystretch{0.8}
\caption[]{Results of the K-S comparison between
data and simulation for the second set of 9 kinematical variables.}
\begin{tabular}{l c   c}
  & \multicolumn{1}{c}{ Events with a superjet} &
    \multicolumn{1}{c}{ Complementary sample} \\
  Variable                   & $P$ (\%)  & $P$ (\%)  \\
  $\MET$                          &  27.1             & 57.1 \\
  $M_T^W$                           &  13.1          & 38.2 \\
  $M^{b+suj}$                      &  4.0          & 58.9 \\
  $y^{b+suj}$                     &  7.1        & 34.9 \\
  $E_T^{b+suj}$                     &  24.0       & 60.1 \\
  $M^{l+b+suj}$                   &  21.0         & 33.6 \\
  $\delta \theta^{b,suj}$          &  30.1        & 41.1 \\
  $\delta \phi^{b,suj}$            &  15.3         & 83.8 \\
  $\delta \theta^{l,b+suj}$       &  37.3       & 35.7 \\
\end{tabular}             
\label{tab:tab_comb.2}
\end{center}
\end{table}

\clearpage
\section{Evaluation of the combined probability}
\label{sec:eval}
 Using the results of the previous section, we first evaluate the combined
 probability that the data are statistically consistent with the simulation
 using the  set of 9 kinematical variables listed in Table~\ref{tab:tab_comb.1}.
 The combined probability is evaluated with three different approaches
 in order to test the sensitivity of the result to the correlations between
 kinematical variables. 

 In the simplest method, we evaluate the probability of observing a value of
 $\Pi ={\displaystyle \prod_{i}^{n}} P_i$, where $n$ is the number of 
 kinematic variables, no larger than that of the data ($\Pi^0$). If the
 kinematical variables are uncorrelated, this probability is
 $\Pi_T= \Pi^0 {\displaystyle \sum_{k=0}^{n-1}} {\displaystyle
 \frac { (-\ln \Pi^0)^{k}} {k !} }$~\cite{sadoulet}.
 This method yields $\Pi_T=0.46$ for the complementary sample and
 $\Pi_T =1.6 \times 10^{-6}$ for events with a superjet.

\begin{figure}[htb]
\begin{center}
\leavevmode
\epsfysize=9.5cm
\epsffile{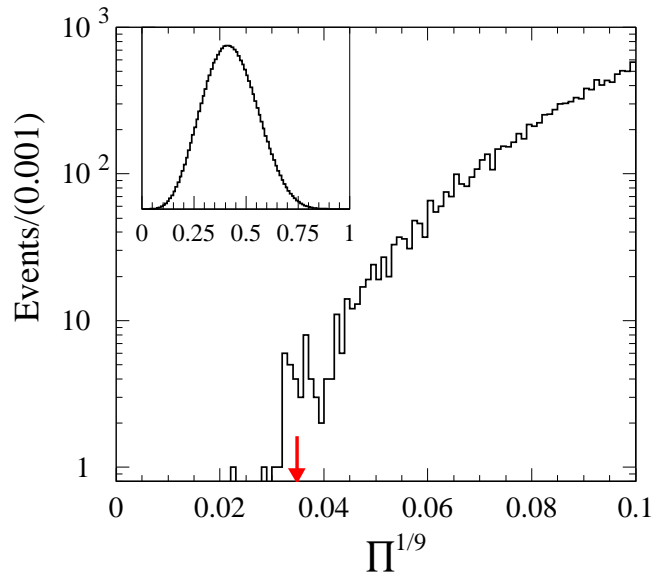}
\caption[]{Distribution of the product $\Pi$ of 9 probabilities obtained with
           $10^{7}$ pseudo-experiments which use 13 events randomly extracted
           from the SM simulation (see text). The arrow indicates the $\Pi$ 
           value of the data. The inset shows the $\Pi$ distribution in full.}
\label{fig:fig_5.12bis}
\end{center}
\end{figure}

 In the second method, which accounts for the effect of correlations between 
 variables, we perform a large number of Monte Carlo pseudo-experiments. In
 each experiment, we form a set of 8 $\W+$ 2 jet and 5 $\W+$ 3 jet different
 events randomly extracted from the simulations of the 12 processes listed in
 Tables~V and~VI of Ref.~\cite{anomal}. In each experiment, we first randomly
 determine $N_i$, the number of events contributed by each process $i$,
 separately for the 2 and 3 jet bin. This is done using as probabilities the 
 ratios $\sigma_i/\sigma$, where the contribution $\sigma_i$ of each process 
 $i$ (as listed in Tables~V and~VI of Ref.~\cite{anomal}) is smeared, in each
 experiment, by its error using a Gaussian distribution and 
 $\sigma= {\displaystyle \sum_{i=1}^{12}} \sigma_i$.  We then randomly  
 extract $N_i$ events from the simulation of each process $i$ to form a sample
 of 13 events (8 with 2 jets and 5 with 3 jets).  We compare the distribution
 of the nine kinematical variables to the SM templates by using the same K-S
 test of Ref.~\cite{anomal} and derive the product $\Pi$ of the probabilities
 $P_i$ for each experiment. The combined probability that the data are 
 consistent with the SM simulation is given by $\Pi_C$, the fraction of
 pseudo-experiments which have a probability $\Pi$ no larger than $\Pi^0$.
 The distribution of the probability product $\Pi$ resulting from $ 10^{7}$ 
 pseudo-experiments which use simulated events is shown in  
 Figure~\ref{fig:fig_5.12bis}. We find 16 pseudo-experiments with a product of
 probabilities  no larger than that observed for the superjet data. This 
 corresponds to a combined probability
 $\Pi_C = (1.6 \pm 0.4) \times 10^{-6}$ (4.8 $\sigma$ effect).

\begin{figure}
\begin{center}
\leavevmode
\epsfysize=9.5cm
\epsffile{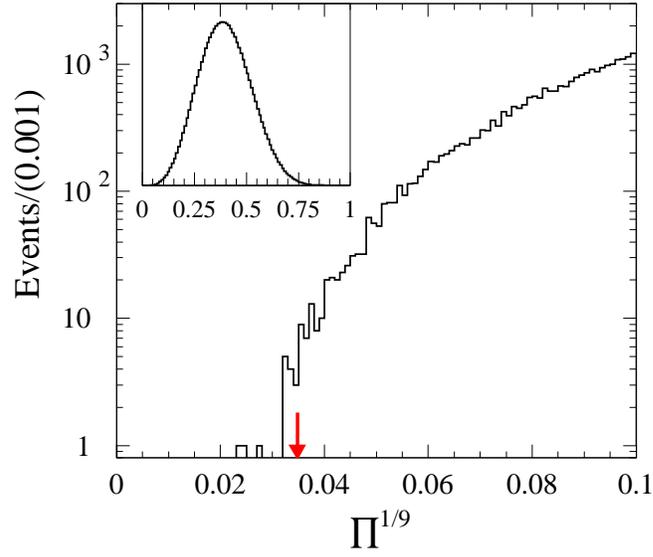}
\caption[]{Distribution of the product $\Pi$ of 9 probabilities of 13 events
           extracted randomly from the complementary sample of 42 events.
           The arrow indicates the $\Pi$ value of the data.}
\label{fig:fig_5.13}
\end{center}
\end{figure}

 We have also performed  pseudo-experiments in which we compare the SM 
 simulation to 13 different events extracted randomly from the complementary
 sample of data consisting of 42 events. For each experiment, we compare the
 kinematical distributions of each sample to the SM templates and derive the
 product of probabilities $\Pi$. Figure~\ref{fig:fig_5.13} shows the
 $\Pi$ distribution of the $10^{7}$ pseudo-experiments. The probability that 13
 events randomly extracted from the control sample have a product $\Pi$
 no larger than the data is $(1.4 \pm 0.4)\times 10^{-6}$. In other words,
 it is very hard to find, among these particular 42 events, a subsample of 
 13 events that disagrees  with the SM simulation as much as the superjet 
 sample.

 We have studied a few  effects which might influence the low value of the 
 combined probability.

 As observed in Section~VD of Ref.~\cite{anomal}, the rapidity
 distributions of the objects in the final state are quite asymmetric.
 Since we know of no physics process that would produce such asymmetries,
 it is possible that they are due to an obscure detector problem, not seen
 in other data samples, or to a low probability statistical fluctuation.
 Therefore, it is of interest to understand the effect of these asymmetries on
 the low value of the combined probability. We have done this by comparing
 the 9 observed and simulated distributions using the pseudo-rapidity absolute
 values. This test also yields a small value of the combined probability 
 ($\Pi_T =4.5 \times 10^{-6}$).

 The combined probability value depends on the estimate of the contribution of
 each SM process and its uncertainty. We have studied the effect of varying
 the fraction of $t\bar{t}$ events. If we make the hypothesis that the data 
 are  contributed only by $t\bar{t}$ events, $\Pi_T$ grows to  
 $1.2 \times 10^{-5}$ for the events with a superjet and decreases to 
 $ 0.8 \times 10^{-2}$ for the complementary sample.

 We next study the bias due to the use of a particular set of kinematical 
 variables which, while quite reasonable and well motivated, was not chosen 
 {\em a priori}. For example, we could have evaluated the combined probability
 using a slightly different set of 8 kinematic variables: $E_T^{l}$, $\eta^l$,
 $E_T^{suj}$, $\eta^{suj}$, $E_T^{b}$, $\eta^{b}$, $\MET$, and $M_{T}^{W}$. 
 This set does not describe the kinematics of all objects in the final state 
 as completely as the previous one, but it contains some variables which are
 more intuitive. In this case, we derive the following combined probabilities:
 $ \Pi_T = 7.4 \times 10^{-5}$ and
 $\Pi_C = (2.5  \pm 0.5) \times 10^{-5}$ (4.2  $\sigma$ effect).
 
 Events with a superjet are not very anomalous when using the set of
 9 kinematical variables listed in Table~\ref{tab:tab_comb.2}. Using this set
 of variables,  the combined probabilities for events with a superjet  are
 $ \Pi_T = 1.9 \times 10^{-2}$ and $\Pi_C = 2.3 \times 10^{-2}$, respectively.

\begin{figure}[htb]
\begin{center}
\leavevmode
\epsfysize=9.5cm
\epsffile{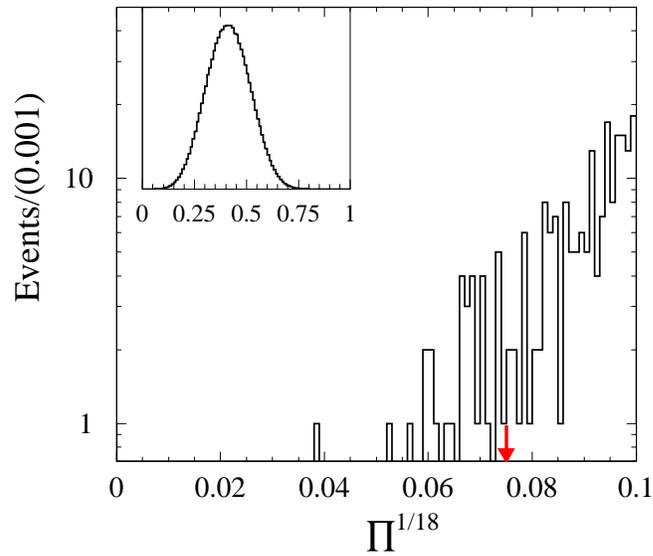}
\caption[]{Distribution of the product $\Pi$ of 18 probabilities obtained with
           $10^{6}$ pseudo-experiments which use 13 events randomly extracted
           from the SM simulation (see text). The arrow indicates the
           $\Pi$ value of the data.}
\label{fig:fig_appb.1}
\end{center}
\end{figure}

 The effect of the bias due to the {\em a posteriori} choice of a particular
 set of kinematical variables is removed by evaluating the combined probability
 for all the 18 kinematical variables inspected. In this case, the probability
 that the data are consistent with the simulation is $\Pi_T = 0.67$
 for the complementary sample and $\Pi_T = 6.0 \times 10^{-7}$ for events with
 a superjet.  This  estimate of the combined probability does not account for
 the effect of large correlations between a few of the 18 kinematical variables.
 With $10^{6}$ pseudo-experiments which use simulated events we evaluate that
 in this case the combined probability for events with a superjet is 
 $\Pi_C=(3.4 \pm 0.6) \times 10^{-5}$. The $\Pi$ distribution resulting from
 these pseudo-experiments is shown in Figure~\ref{fig:fig_appb.1}.
\vspace*{-1.5cm}
\vspace*{-1.5cm}
\vspace*{-1.0cm}
%
 \section{Conclusions}
 \label{sec:s-concl}
 Having taken into account the correlations between kinematical variables, 
 we estimate that the combined probability that the events with a superjet,
 reported by the CDF collaboration in Ref.~\cite{anomal}, are statistically
 consistent with the SM simulation is $(1.6 \pm 0.4) \times 10^{-6}$ 
 (4.8 $\sigma$ effect). This probability is derived using a particular set of
 9 kinematical variables, selected {\em a posteriori} from a larger
 set of 18, which was chosen {\em a priori} in order to search for
 differences between data and simulation. The effect of the bias due to the
 {\em a posteriori} selection of particular sets of variables
 cannot be univocally assessed. We have therefore evaluated the combined
 probability that these events are consistent with the simulation using all
 kinematical variables which have been inspected. We find that the combined 
 probability remains low [$(3.4 \pm 0.6) \times 10^{-5}$ (4.1 $\sigma$ effect)].

\acknowledgments
 We thank the Fermilab staff and the CDF collaboration for their contributions.
 This work was supported by the U.S.~Department of Energy, the National Science
 Foundation and the Istituto Nazionale di Fisica Nucleare.
\end{document}